\documentclass[12pt]{article}
\usepackage{layout,amssymb,amsmath,epsfig}

\begin{document}
\title{\bf Analytic models of Anisotropic Strange Stars in $f(T)$ Gravity
with Off-diagonal tetrad}
\author{M. Zubair$^1$ \thanks{mzubairkk@gmail.com; drmzubair@ciitlahore.edu.pk} and G.
Abbas$^2$ \thanks{ghulamabbas@ciitsahiwal.edu.pk}\\\\
$^1$ Department of Mathematics, COMSATS\\
Institute of Information Technology, Lahore, Pakistan.\\
$^2$ Department of Mathematics, COMSATS\\
Institute of Information Technology, Sahiwal, Pakistan.}

\date{}

\maketitle
\begin{abstract}
This paper is devoted to study the analytic models of anisotropic
compact stars in $f(T)$ gravity (where $T$ is torsion scalar), with
non-diagonal tetrad. By taking the anisotropic source inside the
spherically symmetric star, the equations of motions have been
derived in the context of $f(T)$ gravity. Krori and Barua metric
which satisfies the physical requirement of a realistic star, has
been applied to describe the compact objects like strange stars. We
use the power law form of $f(T)$ model to determine explicit
relations of matter variables. Further, we have found the
anisotropic behavior, energy conditions, stability and surface
redshift of stars. Using the masses and radii of 4$U$1820-30, Her
X-1, SAX J 1808-3658, we have determined the constants involved in
metric components. Finally we discuss the graphical behavior of the
analytic description of strange star candiddates.
\end{abstract}

\section{Introduction}
The accelerated expansion of the Universe \cite{1,2}, and its
verification by several observations \cite{3}-\cite{5}, motivate the
theocratical physicists to propose such theories of gravity, which
can extend General Relativity (GR). The extension should be in such
a way that observation can be justified by the modifications instead
of introducing dark energy. It is well known that GR deals with the
description of gravity at local scale but not at global scale, for
example GR fails to explain the the rotation of spiral galaxies
\cite{6}. Several theories of gravity have proposed to generalized
the Einstein theory of gravity, but most of these theories are the
geometric extension of GR which provide the essential ingredients to
support the observations. Among all these modifications, one can
consider the $f(R)$ theories, in which Lagrangian can be expressed
as a function of Ricci Scalar $R$ \cite{7,8}.

Analogous to $f(R)$ theory of gravity there is another theory
so-called $f(T)$ theory, this theory is defined by the teleparallel
equivalence of gravity (TEGR). In this theory gravitational action
can be described by the torsion scalar $T$ instead of Lagrangian. In
$f(T)$ theory, Riemann-Cartan spacetime is defined with
Weitzzenb$\ddot{o}$ck connections instead of Levi-Civita
connections, in this situation background spacetime admits non-zero
torsion and zero curvature. This definition of spacetime was
introduced by Einstein to define gravitation in terms of torsion and
tetrad. In TEGR, tetrad field plays the role of dynamical field
instead of metric field. Inspite of such differences TEGR and GR
have many similarities, in other words every solution of GR is also
solution of TEGR. But when TEGR is generalized to $f(T)$ gravity
theory, by assuming the Lagrangian as a general function of torsion
scalar, the equivalence between these tow theories breaks down
\cite{11,12}. Therefore $f(T)$ theory can be considered as a
prominent candidate for the explanation of the accelerated expansion
of universe in the absence of dark energy \cite{13}.

In $f(T)$ gravity, equations of motions are second order
differential equations as in GR, but in $f(R)$ such equations are
fourth order in metric formulism (second order in plantini
approach). As compared to GR, dynamical equations in $f(T)$ gravity,
displays the additional degree of freedom which is related to the
fact that equations of motions are invariant under the Lorentz
transformation \cite{15}. Recently \cite{16}, it has been proposed
that there are two types tetrad, bad and good. The diagonal and
non-diapasonal tetrad are known as bad and good tetrad,
respectively. The non-diagonal tetrad are used to properly
parallelize the isotropic spherically symmetric spacetime. No doubt
$f(T)$ theory is excellent theory for the explanation of cosmic
acceleration and observations on large scale(clustering of galaxies)
\cite{17}. But we must remember that GR is in excellent agreement in
with solar system test and pulsar observation \cite{18}. In recent
papers \cite{19,20}, some solar system constraints have been
proposed for the $f(T)$ gravity in the diagonal tetrad case. But in
the present paper, we adopt the non-diagonal tetrad to drive the
models of anisotropic compact stars in $f(T)$, gravity by assuming
the power law form of $f(T)$ model.

In theoretical astrophysics, $f(T)$ theory was used to explore the
effects of $f(T)$ models in 3dimensions, so that $f(T)$ version of
BTZ black hole solutions has been formulated \cite{21}. Later on
\cite{22}, it has been proved that the violation of first of black
hole thermodynamics in $f(T)$ gravity, is due to the violation of
Lorentz invariance. Recently \cite{23}, some static spherically
symmetric solutions with charged source have been found in $f(T)$
theory. After finding a large class of static perfect fluid
solutions \cite{24}, the physical conditions for the existence of
astrophysical stars in $f(T)$ theory have been discussed \cite{25}.

Recently, Abbas and his collaborators have studied the models of
anisotropic compact stars in GR, $f(R)$, $f(G)$ and $f(T)$ theories
in diagonal tetrad case \cite{25a}-\cite{25d}. The main objective of
this paper is to study the symmetric models of the anisotropic
compact stars in the context of $f(T)$ gravity using non-diagonal
tetrad. We investigate the anisotropic the anisotropic behavior,
regularity as well as stability of these models. Finally, the
surface redshift has been calculated. All these properties of the
models have been discussed by using the observational data of the
compact stars. The plan of the paper is the following:  In the next
section, we present the review of $f(T)$ gravity. Section \textbf{3}
deals with anisotropic source and equation of motion equations in
$f(T)$ gravity with nondiagonal tetrad. Section \textbf{4}
investigates with the physical analysis of the proposed models. In
the last section, we conclude the results of the paper.

\section{$f(T)$ Gravity: Fundamentals}

In this section we briefly overview the basics of $f(T)$ gravity
(for review see ). The dynamical variables of TEGR are the
orthonormal tetrad componenets $e_A(x^\mu)$. The metric can be
defined via these components by
$g_{\mu\nu}=\eta_{AB}e^A_\mu{e}^B_\nu$, where
$\eta_{AB}=diag(1,-1,-1,-1)$. Herein, A,B run over $0,1,2,3$ for the
tangent space of the manifold and $\mu,\nu$ are coordinate indices
on the manifold which also run over $0,1,2,3$. The tetrad fields
define a connection of the form
$\Gamma^\lambda_{\mu\nu}=e^\lambda_i\partial_\nu{e}^i_\mu$, which is
named as Weitzenb\"{o}ck's connection. This connection involves no
contribution from curvature as compared to Levi-Civita connection
for which torsion is zero.

The torsion $T^\lambda_{\mu\nu}$ and contorsion $K^{\mu\nu}_\lambda$
tensors are defined by
\begin{eqnarray}\label{1}
T^\lambda_{\mu\nu}=e^\lambda_A(\partial_\mu{e}^A_\nu-\partial_\nu{e}^A_\mu),\\\label{2}
K^{\mu\nu}_\lambda=-\frac{1}{2}(T^{\mu\nu}_\lambda-T^{\nu\mu}_\rho-T^{\mu\nu}_\lambda).
\end{eqnarray}
The torsion scalar defines the teleparallel Lagrangian density and
is given by
\begin{equation}\label{3}
T=T^\lambda_{\mu\nu}S^{\mu\nu}_\lambda,
\end{equation}
where
\begin{equation}\label{4}
S^{\mu\nu}_\lambda=\frac{1}{2}(K^{\mu\nu}_\lambda+\delta^\mu_\lambda{T}^{\alpha\nu}_\alpha
-\delta^\nu_\lambda{T}^{\alpha\mu}_\alpha).
\end{equation}
One can define the modified teleparallel action by replacing $T$
with a function of $T$, in analogy to $f(R)$ gravity \cite{7,8}, as
follows
\begin{equation}\label{5}
\mathcal{I}=\int{dx^4e\left[\frac{1}{2\kappa^2}
f(T)+\mathcal{L}_{(M)}\right]},
\end{equation}
where $e=det(e^A_\mu)=\sqrt{-g}$, $\kappa^2=8\pi{G}=1$. The
variation of the action with respect to the vierbein vector field
$e^\mu_i$ presents
\begin{equation}\label{6}
e^{-1}\partial_\mu(eS^{\mu\nu}_A)f'(T)-e^\lambda_AT^\rho_{\mu\lambda}S^{\nu\mu}_\rho
f'(T)+S^{\nu\mu}_A\partial_\mu(T)f''(T)+\frac{1}{4}e^\nu_Af(T)=\frac{\kappa^2}{2}
e^\rho_AT^{(m)\nu}_\rho,
\end{equation}
where $T^{(m)}_{\mu\nu}$ is the matter energy momentum tensor and
prime denotes the differentiation with respect to $T$.

\subsection{Anisotropic Matter Configuration in $f(T)$ Gravity}

We assume the Krori and Barua spacetime to describes the strange
star stellar configuration, which is given by \cite{26}
\begin{equation}\label{7}
ds^2=-e^{\mu(r)}dt^2+e^{\nu(r)}dr^2+r^2d\Omega^2,
\end{equation}
with $\nu=Ar^2$, $\mu=Br^2+C$ where
$d\Omega^2=d\theta^2+sin^2\theta{d\phi^2}$, $A$, $B$ and $C$ are
arbitrary constant to be evaluated by using some physical
constraints. In this setup we consider the off-diagonal tetrad
matrix of the form \cite{27}
\begin{equation}\label{8}
e^a_\mu=\left(
\begin{array}{cccc}
e^{\mu/2}& 0 & 0 & 0 \\
    0 & e^{\nu/2}sin\theta{cos\phi} & rcos\theta{cos\phi} & -rsin\theta{sin\phi} \\
    0 & e^{\nu/2}sin\theta{sin\phi} & rcos\theta{sin\phi} & rsin\theta{cos\phi} \\
    0 & e^{\nu/2}cos\theta & -rsin\theta & 0 \\
  \end{array}\right)
\end{equation}
The determinant of $e_\mu^i$ can be obtained as
$e=e^{\mu+\nu}r^2sin\theta$.

We assume the anisotropic fluid as interior of compact object
defined by the energy momentum tensor $T^m_{\alpha\beta}$ as
\begin{equation}\label{9}
T^m_{\alpha\beta}=(\rho+p_t)u_\alpha{u}_\beta-p_tg_{\alpha\beta}+(p_r-p_t)v_\alpha{v}_\beta,
\end{equation}
where $u_\alpha=e^\frac{\mu}{2}\delta^0_\alpha$,
$v_\alpha=e^\frac{\nu}{2}\delta^0_\alpha$, $\rho$, $p_r$ and $p_t$
correspond to energy density, radial and transverse pressures,
respectively. The dynamical equations for spacetime (\ref{7}) lead
to
\begin{eqnarray}\label{10}
\rho&=&-\frac{e^{-\nu/2}}{r}(e^{-\nu/2}-1)F'-\left(\frac{T}{2}-\frac{1}{r^2}-
\frac{e^{-\nu}}{r^2}(1-r\nu')\right)\frac{F}{2}+\frac{f}{4},\\\label{11}
p_r&=&\left(\frac{T}{2}-\frac{1}{r^2}+\frac{e^{-\nu}}{r^2}(1+r\mu')\right)
\frac{F}{2}-\frac{f}{4},\\\nonumber
p_t&=&\frac{e^{-\nu}}{2}\left(\frac{\mu'}{2}+\frac{1}{r}-\frac{e^{\nu/2}}{r}\right)F'
+\left(\frac{T}{2}+e^{-\nu}\left(\frac{\nu''}{2}+\left(\frac{\nu'}{4}+\frac{1}{2r}\right)
(\mu'\right.\right.\\\label{12}&-&\left.\left.\nu')\right)\right)\frac{F}{2}-\frac{f}{4},
\end{eqnarray}
where $F=f_T$, prime denotes the derivative with respect to radial
coordinate and $T$ is the torsion scalar given by
\begin{eqnarray}\label{13}
T(r)&=&\frac{2e^{-\nu}(e^{\nu/2}-1)e^{\nu/2}-1-r\mu'}{r^2},
\end{eqnarray}
Eqs.(\ref{10})-(\ref{12}) represents the energy density and
anisotropic pressures for the off-diagonal tetrad. In case of
diagonal tetrad there exist an off-diagonal equation which results
in linear form of algebraic function $f(T)$. However, the choice of
off-diagonal tetrad does not provide any particular condition and
one can set a consistent $f(T)$ model. In this setup we consider a
observationally viable $f(T)$ model given by \cite{28,29}
\begin{equation}\label{14}
f(R)=\beta{T}^n,
\end{equation}
where $\beta$ is an arbitrary constant. In \cite{29} Bamba et al.
discussed the existence of finite-time future singularities for
power law type model of the form $T^{\beta}$, it is found that the
terms involving powers like $\beta>1$ can remove such singularities.
Setare and Darabi \cite{30} developed the $f(T)$ model for power law
solutions and discussed the case when universe enters the phantom
phase.

Using the $f(T)$ model (\ref{14}) together with relations of $\mu$
and $\nu$ in metric (\ref{4}), we find the expressions for $\rho$,
$P_r$ and $P_t$ in the following form
\begin{eqnarray}\nonumber
\rho&=&\frac{1}{(e^{Ar^2/2}-1)(e^{Ar^2/2}-2Br^2-1)^2}\left[2^{n-2}\beta
\left(e^{Ar^2}-(1+2Br^2)^2\right.\right.\\\nonumber&-&\left.\left.4n^2(1+Ar^2+2ABr^4)
+2n(2+3Ar^2+Br^2+6ABr^4+2B^2r^4)\right.\right.\\\nonumber&+&\left.\left.e^{Ar^2}(-3-4n^2-4Br^2
+2n(3+Br^2))+e^{Ar^2/2}(3+8Br^2+2B^2r^4\right.\right.\\\nonumber&+&\left.\left.4n^2(2+Ar^2+ABr^4)
-2n(5+3Ar^2+4Br^2+2ABr^4+2B^2r^4))\right)\right.\\\label{15}&\times&\left.\left(\frac{e^{-Ar^2}
(e^{Ar^2/2}-1)(e^{Ar^2/2}-2Br^2-1)}{r^2}\right)^n\right],\\\nonumber
p_r&=&-\frac{2^{n-2}\beta{e}^{-Ar^2}}{r^2}\left(\frac{e^{-Ar^2}
(e^{Ar^2/2}-1)(e^{Ar^2/2}-2Br^2-1)}{r^2}\right)^{n-1}\left(e^{Ar^2}\right.\\\label{16}&+&\left.
2e^{Ar^2/2}(n-1)(1+Br^2-(2n-1)(1+2Br^2))\right),\\\nonumber
p_t&=&2^{n-4}\beta\left[-4+2n\left(2-\frac{(n-1)Ar^2}{(e^{Ar^2/2}-1)^2}
-\frac{(n-1)r^2(A-4B+2ABr^2)}{(e^{Ar^2/2}-2Br^2-1)^2}
\right.\right.\\\nonumber&+&\left.\left.\frac{An-2B+B(An-B)r^2}{B(e^{Ar^2/2}-1)}-
\frac{2B+(A-4B)n+B(An-B)r^2}{B(e^{Ar^2/2}-2Br^2-1)^2}\right)
\right]\\\label{17}&\times&\left(\frac{e^{-Ar^2}(e^{Ar^2/2}-1)(e^{Ar^2/2}-2Br^2-1)}{r^2}\right)^{n}.
\end{eqnarray}
Using the expressions of energy density, radial and transverse
pressures (\ref{15})-(\ref{17}), we find the corresponding equation
of state (EoS) as follows
\begin{eqnarray}\nonumber
\omega_r&=&\left[(1-e^{Ar^2/2}+2Br^2)(e^{Ar^2}+2e^{Ar^2/2}(n-1)(1+Br^2)
-(2n-1)(1\right.\\\nonumber&+&\left.2Br^2))\right]/\left[e^{3Ar^2/2}-(1+2Br^2)^2
-4n^2(1+Ar^2+2ABr^4)+2n(2\right.\\\nonumber&+&\left.3Ar^2+Br^2+6ABr^4+2B^2r^4)+e^{Ar^2}
(-3-4n^2-4Br^2+2n(3\right.\\\nonumber&+&\left.Br^2))+e^{Ar^2/2}(3+8Br^2+4B^2r^4+4n^2(2+Ar^2+ABr^4)
-2n(5\right.\\\label{18}&+&\left.3Ar^2+4Br^2+2ABr^4+2B^2r^4))\right],\\\nonumber
\omega_t&=&\left[e^{2Ar^2}(n-1)+2e^{3Ar^2/2}(n-1)(n-2-2Br^2)-(1+2Br^2)^2-2n^2
\right.\\\nonumber&\times&\left.(1+Br^2)
(1+A(r^2+2Br^4))+n(3+8Br^2+9B^2r^4+2B^3r^6+A(r^2\right.\\\nonumber&+&\left.3Br^4
+2B^2r^6))+e^{Ar^2}(-2(3+6Br^2+2B^2r^4)-2n^2(3+Br^2+A(r^2\right.\\\nonumber&+&\left.Br^4))
+n(12+16Br^2+5B^2r^4+A(r^2+Br^4)))+2e^{Ar^2/2}(2+6Br^2
\right.\\\nonumber&+&\left.4B^2r^4-n(5+10Br^2+7B^2r^4+B^3r^6+A(r^2+2Br^4))
+n^2(3+2Br^2\right.\\\nonumber&+&\left.Ar^2(2+4Br^2+B^2r^4)))\right]/\left[(e^{Ar^2/2}-1)(e^{3Ar^2/2}
-(1+2Br^2)^2-4n^2(1\right.\\\nonumber&+&\left.Ar^2+2ABr^4)+2n(2+3Ar^2+Br^2+6ABr^4+2B^2r^4)+e^{Ar^2}
(-3\right.\\\nonumber&-&\left.4n^2-4Br^2+2n(3+Br^2))+e^{Ar^2/2}(3+8Br^2+4B^2r^4+4n^2(2+Ar^2
\right.\\\label{19}&+&\left.ABr^4)-2n(5+3Ar^2+4Br^2+2ABr^4+2B^2r^4))\right].
\end{eqnarray}

\section{Physical Analysis}

In this section we explore some features of the anisotropic compact
star which includes anisotropic behavior, matching conditions and
stability constraints.

\subsection{Anisotropic Behavior}

Here, we present the evolution of $\rho$, $p_r$ and $p_t$ as shown
in Figs.1-3 for the strange star candidates Her X-1, SAX J
1808.4-3658 and 4U 1820-30 (see table 1). Herein, we set $n=2$ and
$\beta=-2$.
\begin{figure}
\centering \epsfig{file=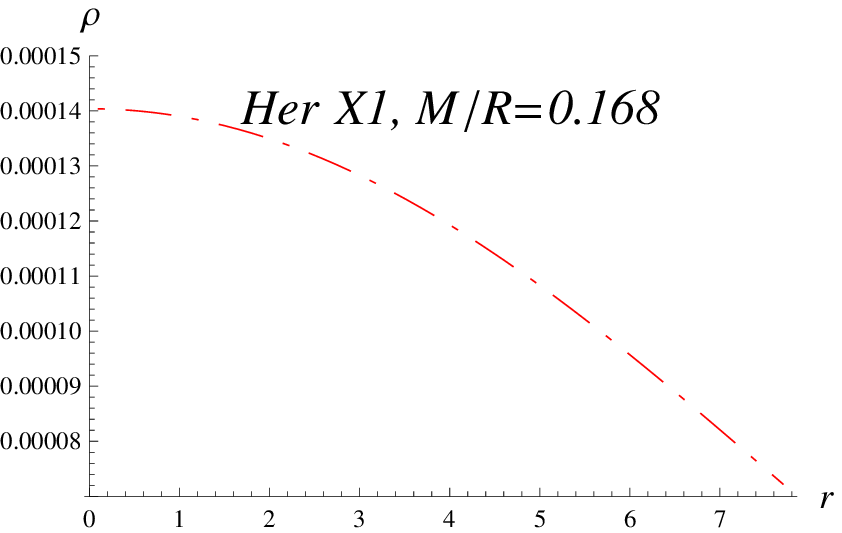, width=.34\linewidth,
height=1.3in}\epsfig{file=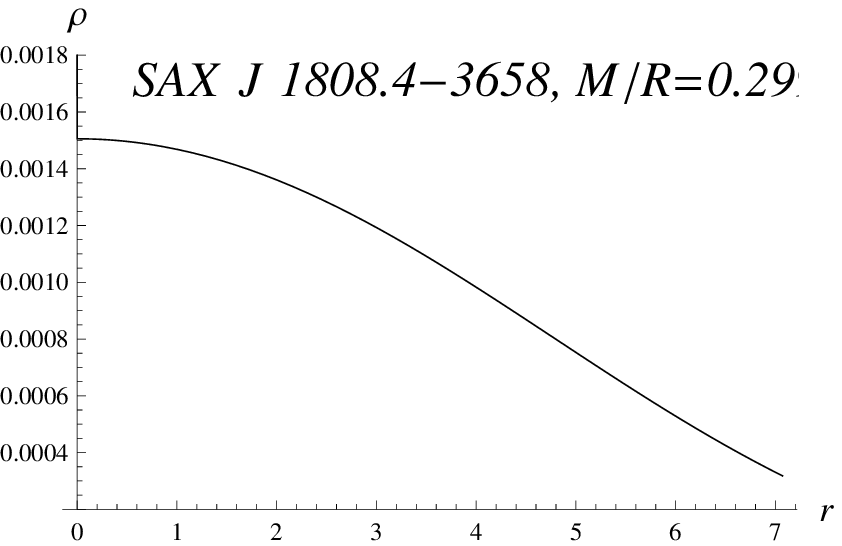, width=.36\linewidth,
height=1.3in}\epsfig{file=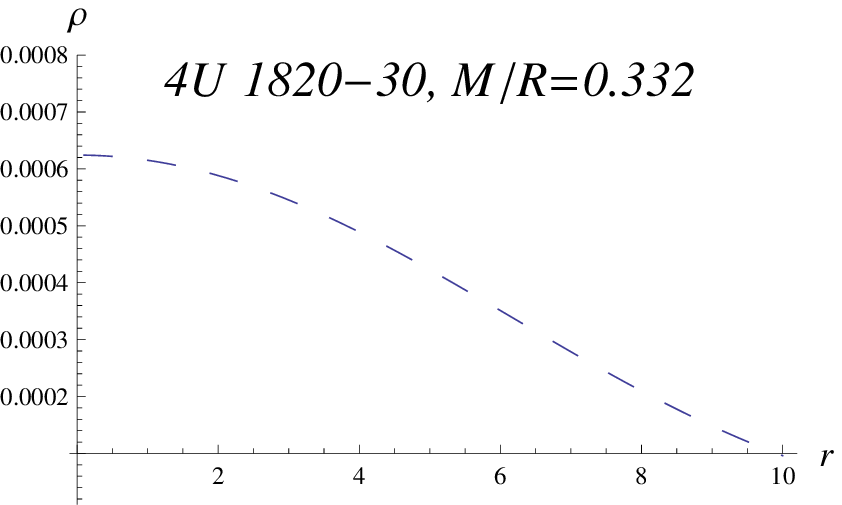, width=.34\linewidth,
height=1.3in}\caption{Evolution of energy density $\rho$ versus
radial coordinate $r(km)$ at the stellar interior of strange star
candidates.}
\end{figure}
\begin{figure}
\centering \epsfig{file=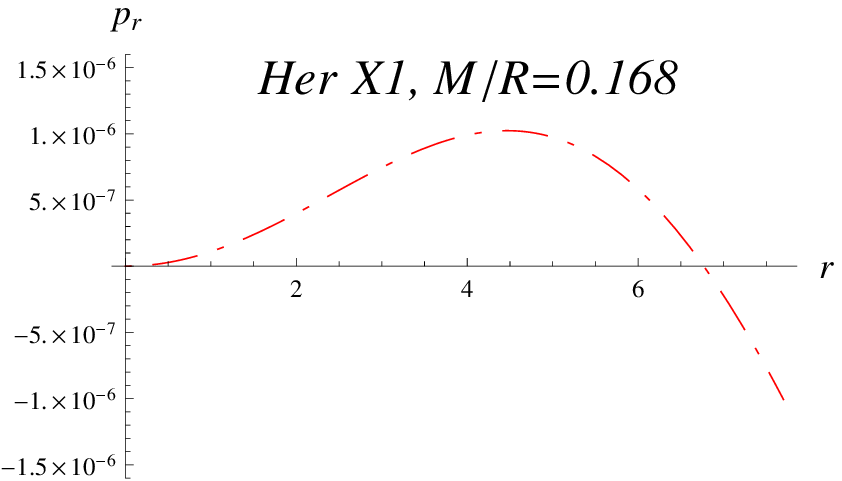, width=.34\linewidth,
height=1.3in}\epsfig{file=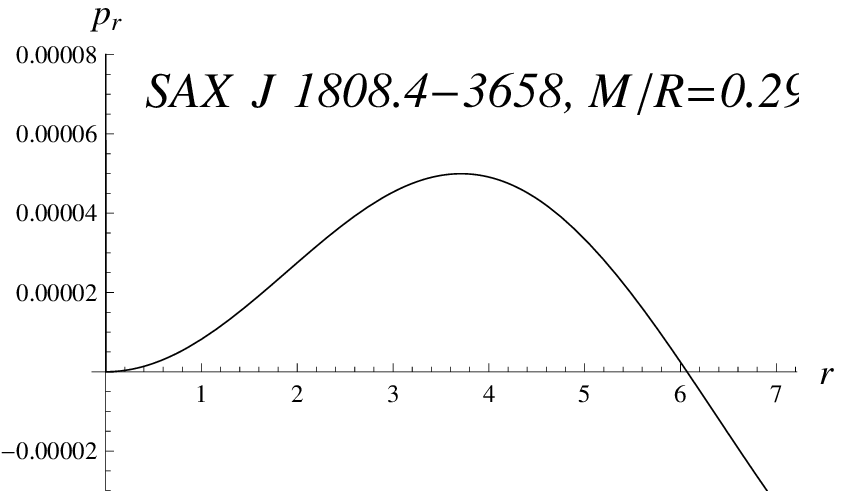, width=.36\linewidth,
height=1.3in}\epsfig{file=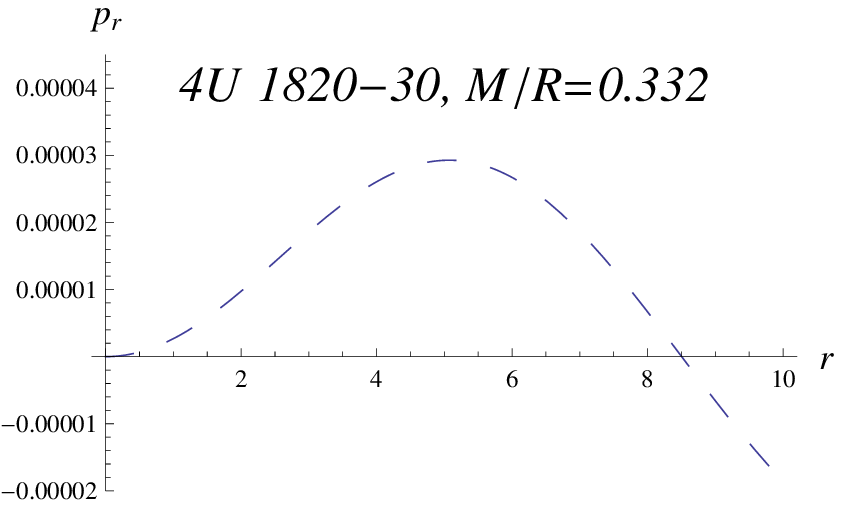, width=.34\linewidth,
height=1.3in}\caption{Evolution of radial pressure $p_r$ versus
radial coordinate $r(km)$ at the stellar interior of strange star
candidates.}
\end{figure}
\begin{figure}
\centering \epsfig{file=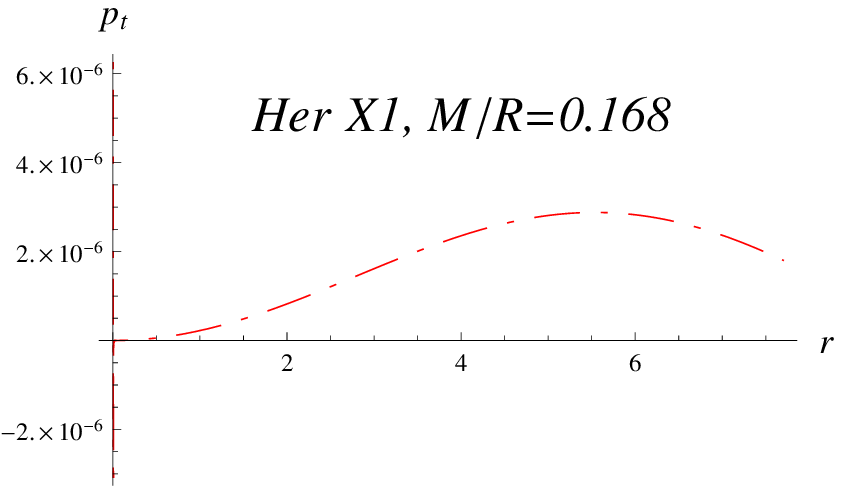, width=.34\linewidth,
height=1.3in}\epsfig{file=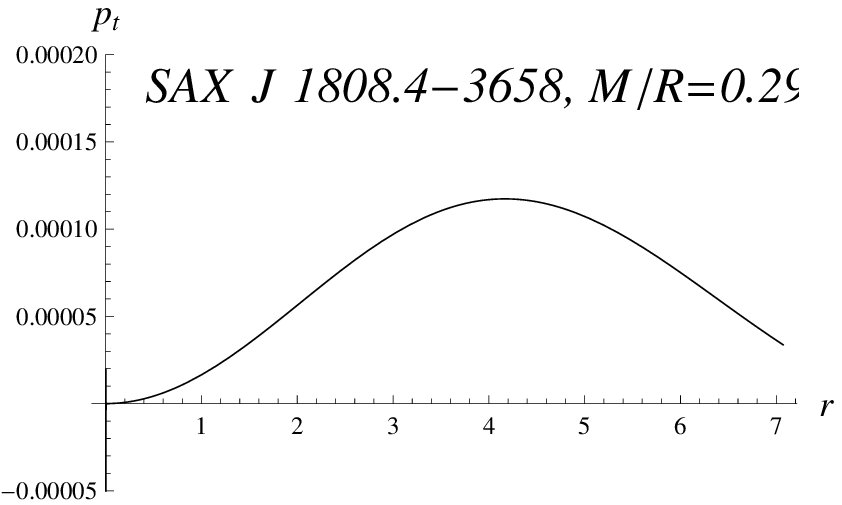, width=.36\linewidth,
height=1.3in}\epsfig{file=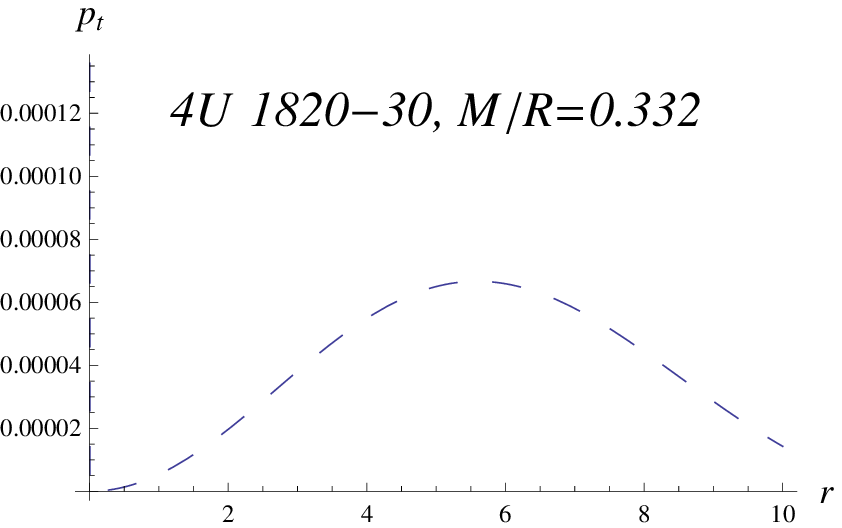, width=.34\linewidth,
height=1.3in}\caption{Evolution of tangential pressure $p_t$ versus
radial coordinate $r(km)$ at the stellar interior of strange star
candidates.}
\end{figure}
From Eqs.(\ref{15}) and (\ref{16}), one can find the expressions for
$d\rho/dr$, $d^2\rho/dr^2$, $dp_r/dr$ and $d^2\rho/dr^2$. We show
the results for $d\rho/dr$ and $dp_r/dr$ in \textbf{Appendix} as
Eqs.(\ref{01}) and (\ref{02}) respectively. Figures 4 and 5 show the
variation of derivatives of $\rho$ and $p_r$ with respect to radial
coordinate for strange star Her X1. We find that $d\rho/dr<0$ and
$dp_r/dr$ shows transition from positive to negative values.
\begin{figure}
\centering \epsfig{file=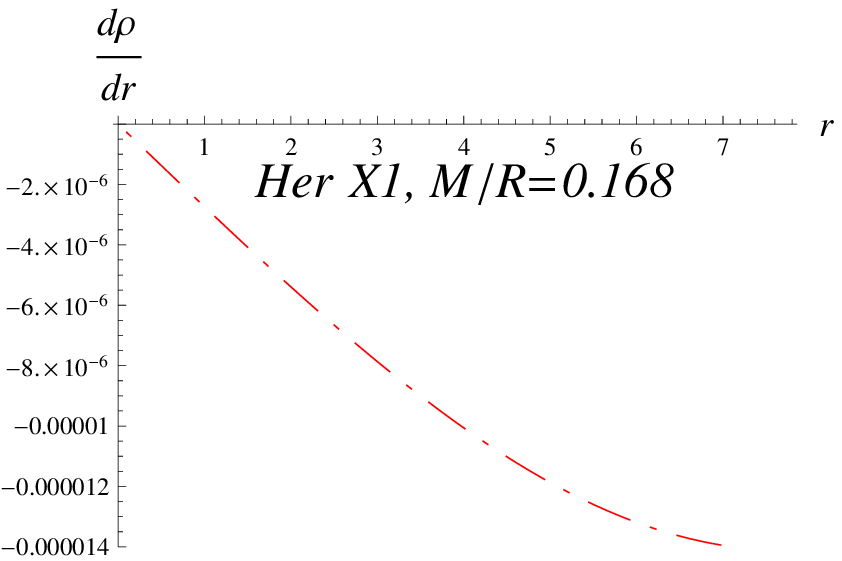, width=.49\linewidth,
height=1.5in}\epsfig{file=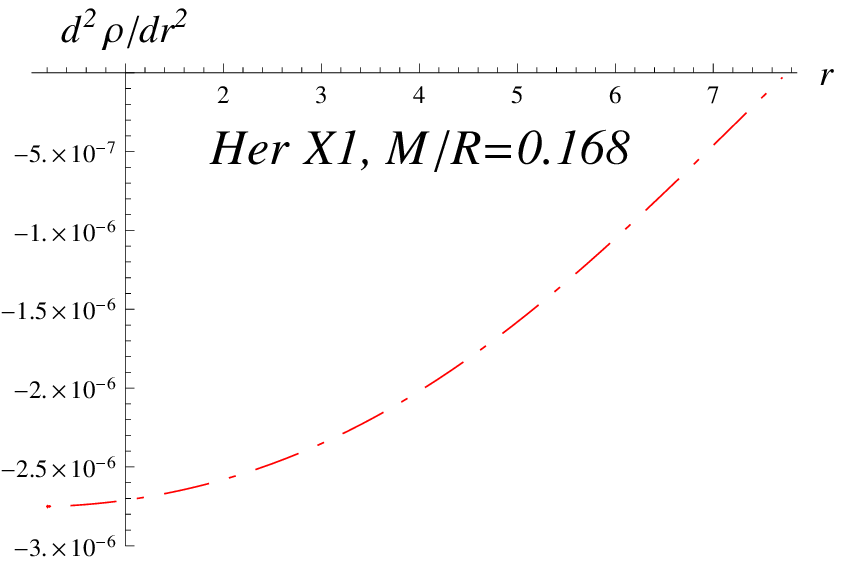, width=.49\linewidth,
height=1.45in}\caption{Evolution of $d\rho/dr$ and $d^2\rho/dr^2$
versus $r(km)$ at the stellar interior of Her X1.}
\end{figure}
\begin{figure}
\centering \epsfig{file=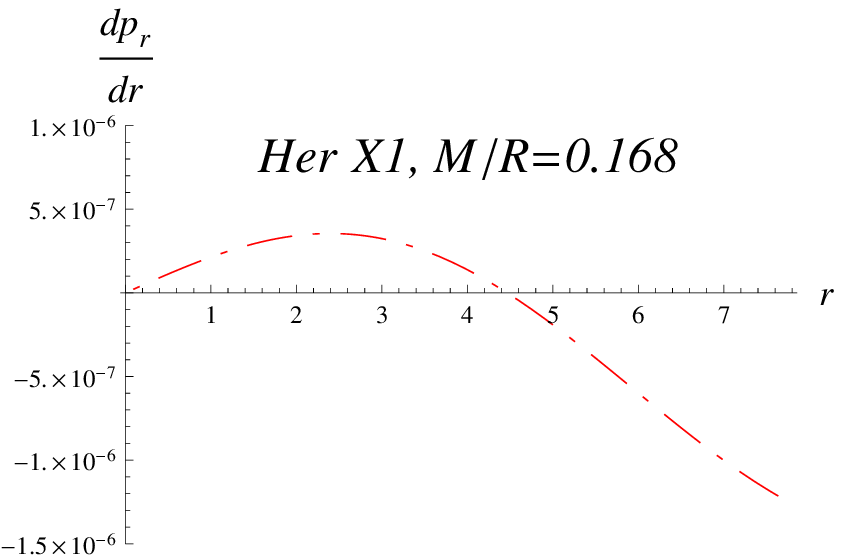, width=.49\linewidth,
height=1.5in}\epsfig{file=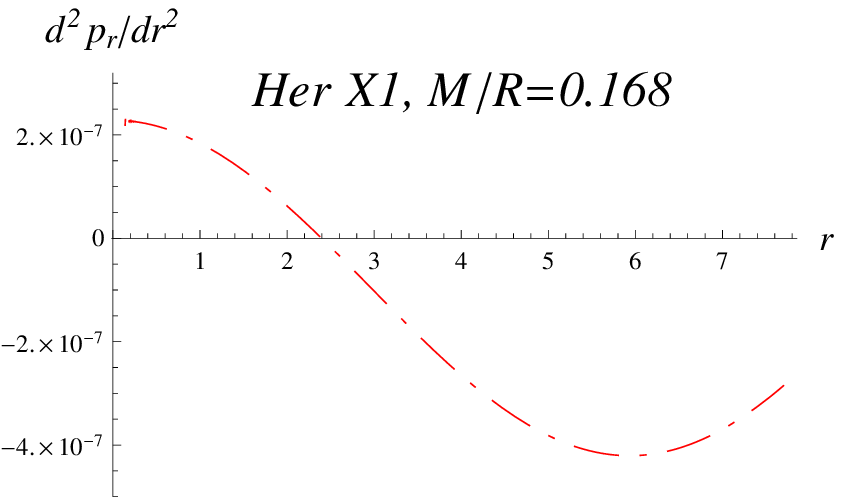, width=.49\linewidth,
height=1.5in}\caption{Evolution of $dp_r/dr$ and $d^2p_r/dr^2$
versus $r(km)$ at the stellar interior of Her X1.}
\end{figure}

We also examine behavior of derivatives at $r=0$, centre of strange
star Her X1. It can be seen that
\begin{eqnarray}\nonumber
\frac{d\rho}{dr}=0, \quad \quad \frac{dp_r}{dr}=0,\\\label{20}
\frac{d^2\rho}{dr^2}<0, \quad \quad \frac{d^2p_r}{dr^2}>0.
\end{eqnarray}
Clearly, at $r=0$, density is maximum and it decreases outward with
the increase in $r$ as shown in Figure 1. In case of $p_r$, we have
increasing function unlike the evolution $\rho$, which attains
maximum value in midway to the boundary of star and then decreases.
One can test the evolution of $\rho$ and $p_r$ for the strange stars
SAX J 1808.4-3658 and 4U 1820-30, which would show similar behavior.
The evolution of EoS parameters $\omega_r$ and $\omega_t$ is also
presented for the strange stars in $f(T)$ gravity. It can be seen
that $\omega_r>0$, which favors the quintessence regime near the
boundary of star whereas $0<\omega_t<1$ despite of the role of
modified gravity.
\begin{figure}
\centering \epsfig{file=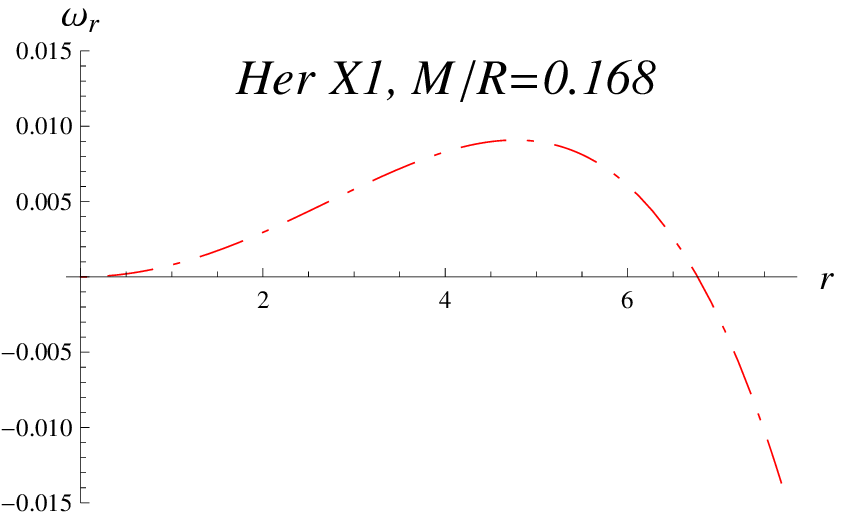, width=.34\linewidth,
height=1.3in}\epsfig{file=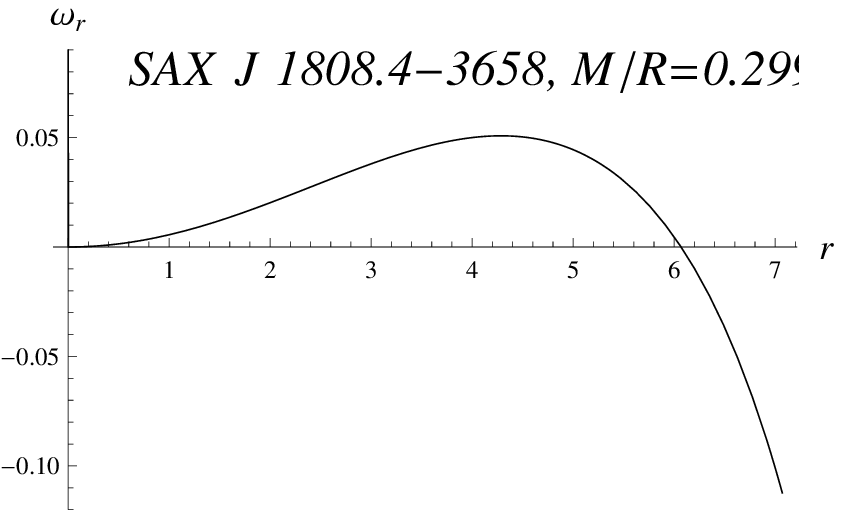, width=.36\linewidth,
height=1.3in}\epsfig{file=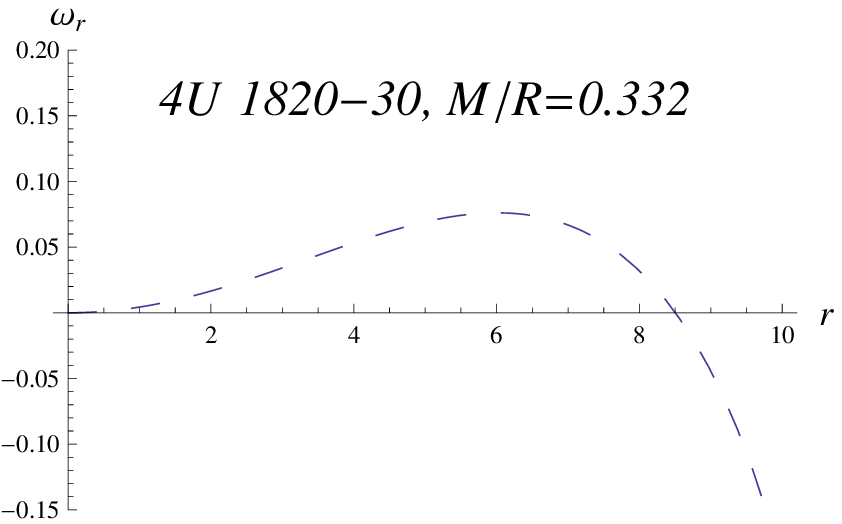, width=.34\linewidth,
height=1.3in}\caption{Evolution of $\omega_r$ versus radial
coordinate $r(km)$ at the stellar interior of strange star
candidates.}
\end{figure}\\
\begin{figure}
\centering \epsfig{file=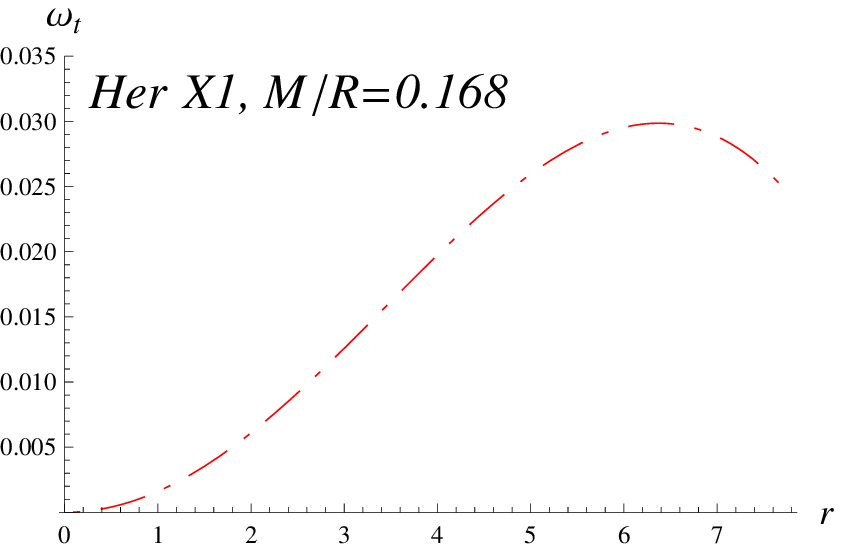, width=.34\linewidth,
height=1.3in}\epsfig{file=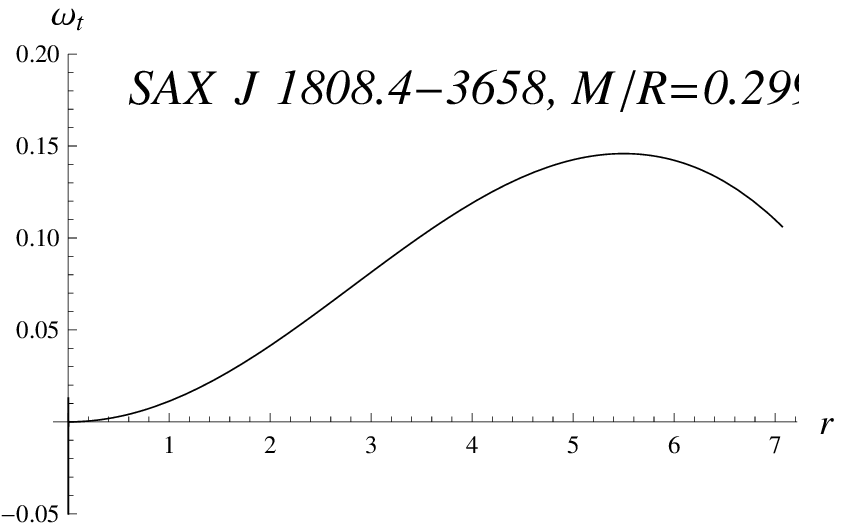, width=.36\linewidth,
height=1.3in}\epsfig{file=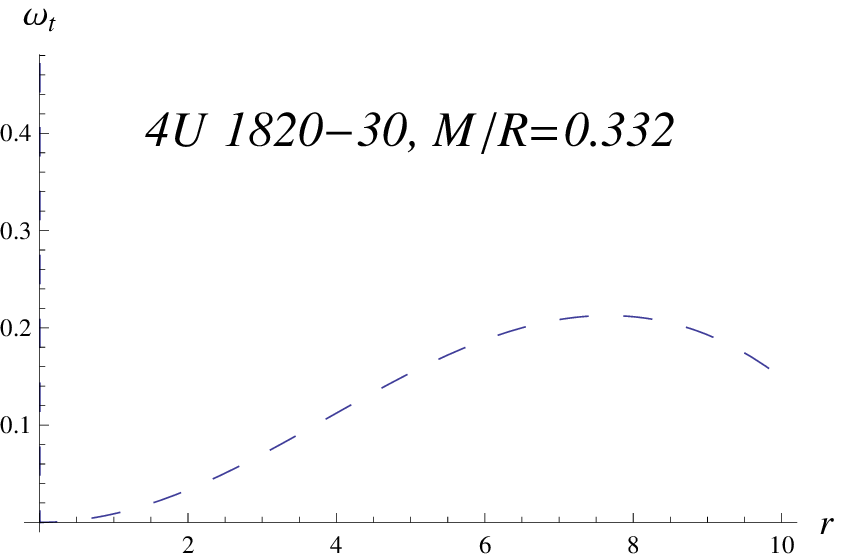, width=.34\linewidth,
height=1.3in}\caption{Evolution of radial pressure $\omega_t$ versus
radial coordinate $r(km)$ at the stellar interior of strange star
candidates.}
\end{figure}

The anisotropy measurement in $f(T)$ gravity, for spherically
symmetric metric is given by
\begin{eqnarray}\nonumber
\Delta&=&\frac{2(p_t-p_r)}{r}=\frac{n2^{n-1}\beta}{r^{2n+1}}
\left\{(e^{-Ar^2})^n\left(\left(e^{\frac{Ar^2}{2}}-1\right)
\left(e^{\frac{Ar^2}{2}}-1-2Br^2\right)\right)^{n-2}\right\}
\\\nonumber&\times&\left[e^{2Ar^2}+2e^{\frac{3Ar^2}{2}}(-2+n-Br^2)+e^{Ar^2}
(6+Ar^2+4Br^2+ABr^4\right.\\\nonumber&&\left.+B^2r^4-2n(3+Ar^2+Br^2+ABr^4))
+(1+Br^2)(1+Ar^2-Br^2\right.\\\nonumber&&\left.+2ABr^4+2B^2r^4-2n(1
+Ar^2+2ABr^4))+2e^{\frac{Ar^2}{2}}(-2-Ar^2-Br^2\right.\\\label{21}&&\left.-2ABr^4-B^2r^4-B^3r^6+n(3
+2Br^2+Ar^2(2+4Br^2+B^2r^4)))\right].
\end{eqnarray}
The measure of anisotropy depends on radial and tangential
pressures, it is directed outward when $p_t>p_r$ which implies
$\Delta>0$ whereas $p_t<p_r$ results in $\Delta<0$ so that
anisotropy is directed inward. We present the evolution of $\Delta$
for strange star candidates Her X-1, SAX J 1808.4-3658 and 4U
1820-30 in Fig. 8. It can be seen that $\Delta>0$ for our model
which implies the existence of repulsive force, allowing more
massive distribution. One can see that $f(T)$ gravity does not
affect the anisotropic force. Here, $\Delta$ vanishes at the center
$r=0$.
\begin{figure}
\centering \epsfig{file=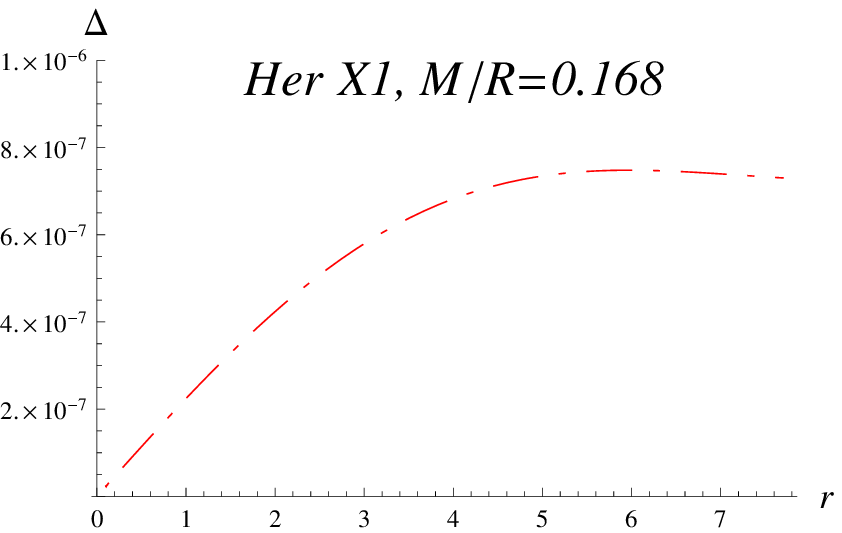, width=.34\linewidth,
height=1.3in}\epsfig{file=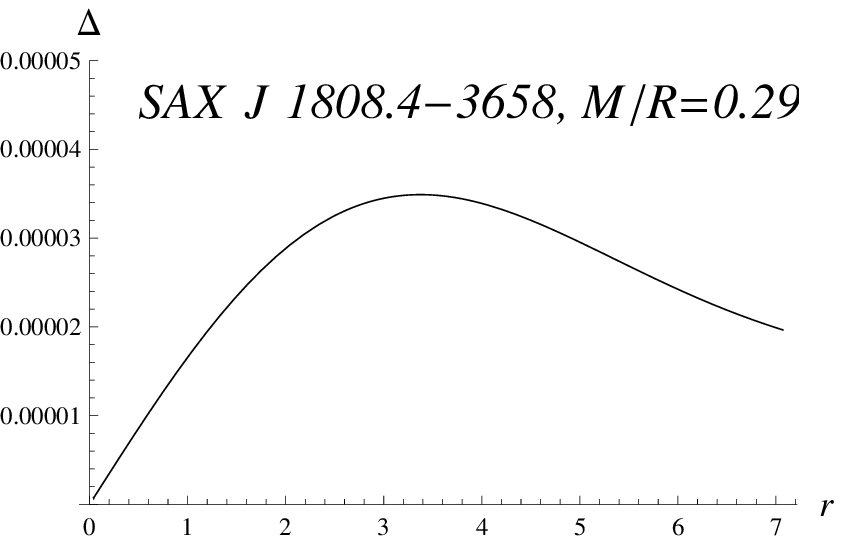, width=.36\linewidth,
height=1.3in}\epsfig{file=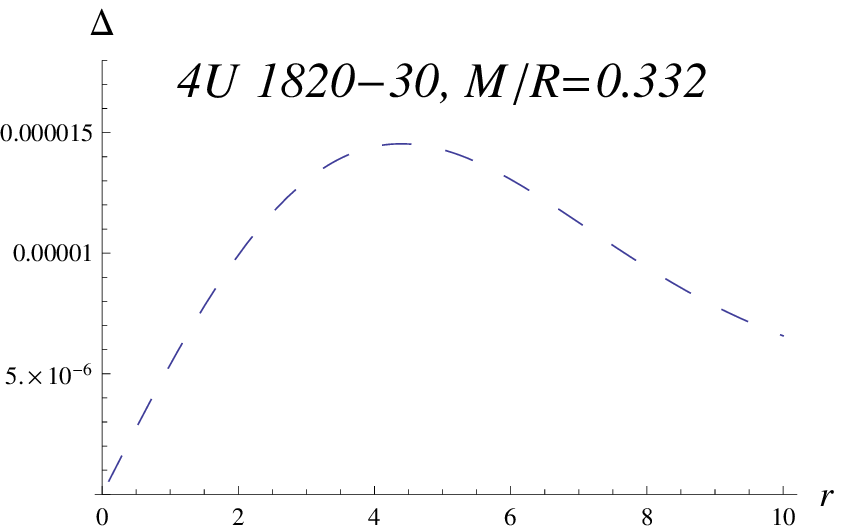, width=.34\linewidth,
height=1.3in}\caption{Evolution of anisotropy measurement $\Delta$
versus radial coordinate $r(km)$ at the stellar interior of strange
star candidates.}
\end{figure}

\subsection{Matching Conditions}

Here, we match the interior metric (\ref{7}) to the vacuum exterior
spherically symmetric metric given by
\begin{equation}\label{22}
 ds^2=-\left(1-\frac{2M}{r}\right)dt^2+\left(1-
 \frac{2M}{r}\right)^{-1}dr^2+r^2d\theta^2+r^2Sin^2\theta{d}\varphi^2,
\end{equation}
At the boundary surface $r=R$ continuity of the metric functions
$g_{tt}$, $g_{rr}$ and $\frac{\partial g_{tt}}{\partial r}$ yield,
\begin{eqnarray}\label{23}
  g_{tt}^-=g_{tt}^+,~~~~~
   g_{rr}^-=g_{rr}^+,~~~~~
   \frac{\partial g_{tt}^-}{\partial r}=\frac{\partial g_{tt}^+}{\partial r},
  \end{eqnarray}
where $-$ and $+$, correspond to interior and exterior solutions.
From the interior and exterior metrics, we get
\begin{eqnarray}\label{24}
  A&=&-\frac{1}{R^2}ln\left(1-\frac{2M}{R}\right),\\\label{25}
 B&=&\frac{M}{R^3}{{\left(1-\frac{2M}{R}\right)}^{-1}},\\\label{26}
 C&=&ln\left(1-\frac{2M}{R}\right)-\frac{M}{R}{{\left(1-\frac{2M}{R}\right)}^{-1}}.
\end{eqnarray}
For the given values of $M$ and $R$ for given star, the constants
$A$ and $B$ takes the values as given in the table \textbf{1}.
\begin{table}[ht]
\caption{Values of constants for given Masses and Radii of Stars
\cite{33a, 33b}}
\begin{center}
\begin{tabular}{|c|c|c|c|c|c|}
\hline {Strange Quark Star}&  \textbf{ $M$} & \textbf{$R(km)$} &
\textbf{ $\frac{M}{R}$} &\textbf{ $A(km ^{-2})$}& \textbf{$B(km
^{-2})$}
\\\hline  Her X-1& 0.88$M_\odot$& 7.7&0.168&0.006906276428 &
$0.004267364618$
\\\hline SAX J 1808.4-3658& 1.435$M_\odot$& 7.07&0.299& 0.01823156974 &
$0.01488011569$
\\\hline 4U 1820-30&2.25$M_\odot$& 10.0 &0.332&0.01090644119 &
$0.009880952381$
\\\hline
\end{tabular}
\end{center}
\end{table}

\subsection{Energy Conditions}

Energy conditions have been a handy tool to limit the arbitrariness
in the energy-momentum tensor, based on Raychaudhuri equation with
attractiveness property of gravity. These conditions include weak
energy condition (WEC), null energy condition (NEC), strong energy
condition (SEC) and dominant energy condition (DEC). In case of
anisotropic fluid, we find the following inequalities
\begin{eqnarray}\nonumber
\textbf{NEC}:\quad&&\rho+p_r\geq0, \quad \rho+p_t\geq0,\\\nonumber
\textbf{WEC}:\quad&&\rho\geq0, \quad \rho+p_r\geq0, \quad
\rho+p_t\geq0,\\\nonumber \textbf{SEC}:\quad&&\rho+p_r\geq0, \quad
\rho+p_t\geq0, \quad \rho+p_r+2p_t\geq0,\\\nonumber
\textbf{DEC}:\quad&&\rho>|p_r|, \quad \rho>|p_t|.
\end{eqnarray}
We examine these constraints for the strange star Her X-1 and show
the evolution in Figure 9. It can be seen that these conditions are
satisfied for strange star Her X-1 and one can also the constraints
on other strange star models.
\begin{figure}
\centering \epsfig{file=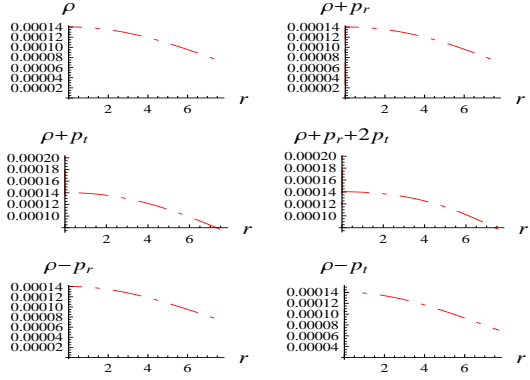, width=.52\linewidth,
height=2in}\caption{Evolution of energy constraints at the stellar
interior of strange star Her X-1.}
\end{figure}

\subsection{TOV Equation}

For an anisotropic fluid the generalized Tolman-Oppenheimer-Volkoff
(TOV) equation has the following form
\begin{equation}\label{27}
\frac{dp_r}{dr}+\frac{\nu'(\rho+p_r)}{2}+\frac{2(p_r-p_t)}{r}=0.
\end{equation}
Following \cite{28a}, we can express the above equation in terms of
gravitational mass and henceforth it results in the equilibrium
condition for the strange star, involving the gravitational,
hydrostatic and anisotropic forces of the stellar object as
\begin{eqnarray}\nonumber
&&F_g+F_h+F_a=0, \\\label{28} && F_g=-Br(\rho+p_r),\quad
F_h=-\frac{dp_r}{dR}, \quad F_a=\frac{2(p_t-p_r)}{r}.
\end{eqnarray}
Using the effective $\rho$, $p_r$ and $p_t$ (\ref{10})-(\ref{12}),
we can find the numerically the equilibrium condition in $f(T)$
theory. In Figure 10, we show the evolution of above forces at the
interior of strange star Her X-1. This figure indicates that the
static equilibrium can be attained due to pressure anisotropy,
gravitational and hydrostatic forces for strange star candidates in
$f(T)$ gravity.
\begin{figure}
\centering \epsfig{file=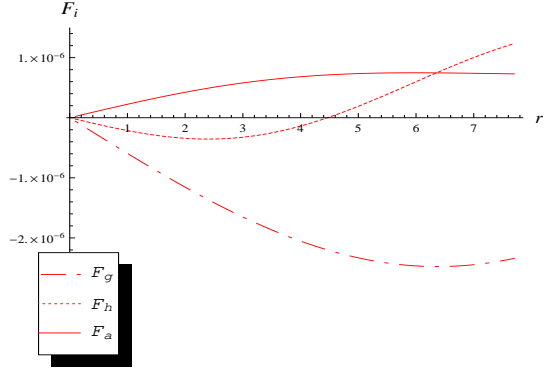, width=.52\linewidth,
height=2in}\caption{Evolution of gravitating, hydrostatic and
pressure anisotropic forces at the stellar interior of strange star
Her X-1.}
\end{figure}

\subsection{Stability Analysis}

In this section, we discuss the stability of strange star models in
$f(T)$ theory. The radial and transverse speeds can be calculated as
\begin{eqnarray}\nonumber
&&v_{sr}^2=\left[(-1+ e^{\frac{Ar^2}{2}}- 2 B r^2) (e^{2 A r^2} +2
e^{\frac{3Ar^2}{2}} (n-2) (1 + B r^2) - (2n-1) (1
\right.\\\nonumber&&\left.+ 2 B r^2) (1 +A (r^2 + 2 B r^4)) + e^(A
r^2) (6 + 10 B r^2 + A (r^2 + 2 B r^4 + 2 B^2 r^6)
\right.\\\nonumber&&\left.- 2 n (3 + 4 B r^2+ A (r + B r^3)^2)) + 2
e^{\frac{Ar^2}{2}} (-(1 + 2 B r^2) (2 + A (r^2 + B r^4))
\right.\\\nonumber&&\left.+ n (3 + 5 B r^2 + 2 A (3 B r^4 +r^2+ 2
B^2 r^6))))\right]/\left[1 -e^{\frac{5Ar^2}{2}} + 7 A r^2 - 10 B r^2
\right.\\\nonumber&&\left.+ 18 A B r^4 + 8 B^2 r^4 + 12 A B^2 r^6
 + 8 A B^3 r^8 +e^{2 A r^2} (5 + 4 n^2 - 2
A r^2 + 6 B r^2 \right.\\\nonumber&&\left.+ 2 n (-3 + A r^2 - B
r^2)) + 4 n^2 (1 + A (r^2+ 2 B r^4))^2 - 2 n (2 - 7 B r^2 + 2 B^2
r^4 \right.\\\nonumber&&\left.+ 3 r^4 (A + 2 A B r^2)^2 +A r^2 (7 +
15 B r^2 + 4 B^2 r^4 + 4 B^3 r^6)) -e^{\frac{3Ar^2}{2}} (10 + 4 B
r^2 \right.\\\nonumber&&\left.+ 8 B^2 r^4 - 2 A^2 r^4 (3 + 2 B r^2)
+A r^2 (16 B r^2 + 2 B^2 r^4 +3) + 8 n^2 (2 + A (r^2
\right.\\\nonumber&&\left.+ B r^4)) +2 n (-11 + 2 B r^2 - 2 B^2 r^4
-A r^2 (5 + 12 B r^2 + B^2 r^4) + 2 A^2 (r^4
\right.\\\nonumber&&\left.+ B r^6))) +e^{A r^2}(-4 A^2 r^4 (3 + 5 B
r^2) +2 (5 - 10 B r^2 + 8 B^2 r^4) +A r^2 (19
\right.\\\nonumber&&\left.+ 56 B r^2 + 16 B^2 r^4 + 4 B^3 r^6) + 4
n^2 (6 + r^4 (A + A B r^2)^2 + A (6 r^2 + 8 B r^4))
\right.\\\nonumber&&\left.-2 n (15 - 14 B r^2 + 6 B^2 r^4 +A^2 r^4
(-1 - 2 B r^2 + 2 B^2 r^4) + 2 A r^2 ( 21 B r^2
\right.\\\nonumber&&\left.+ 4 B^2 r^4 + B^3 r^6 +10))) +
e^{\frac{Ar^2}{2}} (-5 + 28 B r^2 - 16 B^2 r^4 +2 A^2 r^4 (3 + 8 B
r^2 \right.\\\nonumber&&\left.+ 4 B^2 r^4) -A r^2 (21 + 58 B r^2 +
30 B^2 r^4 + 12 B^3 r^6) -8 n^2 (2 + A r^2 (3 + 5 B r^2)
\right.\\\nonumber&&\left.+ A^2 (r^4 + 3 B r^6 + 2 B^2 r^8)) +2 n (9
- 18 B r^2 + 6 B^2 r^4 +A r^2 (21 + 45 B r^2
\right.\\\label{29}&&\left.+ 13 B^2 r^4 + 6 B^3 r^6) +4 A^2 (r^4 + 3
B r^6 + 2 B^2 r^8)))\right].
\end{eqnarray}
\begin{eqnarray}\nonumber&&
v_{st}^2=-(-1 + e^{\frac{3Ar^2}{2}} (n-1) - 2 A r^2 - 10 A B r^4 - 6
B^2 r^4 - 18 A B^2 r^6 - 6 B^3 r^6 \\\nonumber&&- 12 A B^3 r^8 - 4
B^4 r^8 +    e^{\frac{5Ar^2}{2}} (n-1) (-6 + 2 n - B r^2 (4 + A
r^2))-2 n^2 (1 \\\nonumber&&+ B r^2) (1 + A (r^2 + 2 B r^4))^2 + n(3
+ 2 B r^2 + 5 B^2 r^4 + 2 B^3 r^6 +r^4 (1 + B
r^2)\\\nonumber&&\times(A + 2 A B r^2)^2 +A r^2 (6 + 25 B r^2 + 39
B^2 r^4 +28 B^3 r^6 + 4 B^4 r^8)) - e^{2 A r^2} (15 \\\nonumber&&+
16 B r^2 + 6 B^2 r^4 +A r^2 (2 + 10 B r^2 + 3 B^2 r^4) - A^2 (r^4 +
B r^6) +2 n^2 (5 + B r^2\\\nonumber&&+ 2 A (r^2 + B r^4)) +n (-25 -
18 B r^2 - 5 B^2 r^4 -A r^2 (6 + 17 B r^2 + 4 B^2 r^4) + 2 A^2
(r^4\\\nonumber&& + B r^6))) + e^{\frac{3Ar^2}{2}} (20 + 24 B r^2 +
24 B^2 r^4 + 6 B^3 r^6 - 3 A^2 (r^4 + 2 B r^6) + A r^2 (34 B
r^2\\\nonumber&&+8 + 15 B^2 r^4 + B^3 r^6) + 2 n^2 (10 + 4 B r^2 +
r^4 (A + A B r^2)^2 + 2 A r^2 (6 B r^2 + B^2 r^4\\\nonumber&&+4)) +
n (A^2 r^4 (5 + 10 B r^2 + 2 B^2 r^4) - 2 (20 + 16 B r^2 + 10 B^2
r^4 + B^3 r^6) - A r^2 (24 \\\nonumber&&+ 70 B r^2 + 35 B^2 r^4 + 5
B^3 r^6))) - e^{A r^2} (15 + 16 B r^2 + 36 B^2 r^4 + 18 B^3 r^6 + 4
B^4 r^8\\\nonumber&& - 3 A^2 (r^4 + 3 B r^6 + 2 B^2 r^8) + A r^2 (12
+ 52 B r^2 + 39 B^2 r^4 + 6 B^3 r^6 - 2 B^4 r^8)\\\nonumber&& + 2
n^2 (10 + 6 B r^2 + 4 A r^2 (3 + 6 B r^2 + 2 B^2 r^4) + A^2 r^4 (3 +
9 B r^2 + 7 B^2 r^4 + B^3 r^6)) +\\\nonumber&& n (-35 - 28 B r^2 -
30 B^2 r^4 - 6 B^3 r^6 + A^2 r^4 (3 + 9 B r^2 + 8 B^2 r^4 + 2 B^3
r^6) - A r^2 (36
\\\nonumber&&+ 118 B r^2 + 97 B^2 r^4 + 30 B^3 r^6 + 2 B^4 r^8))) +
e^{\frac{Ar^2}{2}} (6 + 4 B r^2 + 24 B^2 r^4 + 18 B^3 r^6
\\\nonumber&&+ 8 B^4 r^8 + A r^2 (8 + 37 B r^2 + 45 B^2 r^4 + 17 B^3
r^6 - 2 B^4 r^8) - A^2 (r^4 + 4 B r^6 + 6 B^2 r^8\\\nonumber&& + 4
B^3 r^10) + 2 n^2 (5 + 4 B r^2 + 2 A r^2 (4 + 10 B r^2 + 5 B^2 r^4)
+ A^2 r^4 (12 B r^2 + 14 B^2 r^4\\\nonumber&&+3 + 4 B^3 r^6)) + n
(-2 (8 + 6 B r^2 + 10 B^2 r^4 + 3 B^3 r^6) + A^2 r^4 (-1 - 4 B r^2 -
2 B^2 r^4
\\\nonumber&&+ 4 B^3 r^6) - A r^2 (24 + 89 B r^2 + 105 B^2 r^4 + 53
B^3 r^6 + 6 B^4 r^8))))/((-1 + e^{\frac{Ar^2}{2}} (1
\\\nonumber&&- e^{\frac{5Ar^2}{2}} + 7 A r^2 - 10 B r^2 + 18 A B r^4
+ 8 B^2 r^4 + 12 A B^2 r^6 + 8 A B^3 r^8 + e^{2 A r^2} (5 + 4
n^2\\\nonumber&& - 2 A r^2 + 6 B r^2 + 2 n (-3 + A r^2 - B r^2)) + 4
n^2 (1 + A (r^2 + 2 B r^4))^2 - 2 n (2 - 7 B r^2\\\nonumber&& + 2
B^2 r^4 + 3 r^4 (A + 2 A B r^2)^2 + A r^2 (7 + 15 B r^2+ 4 B^2 r^4 +
4 B^3 r^6)) - e^{\frac{3Ar^2}{2}}(10
\\\nonumber&&+ 4 B r^2 + 8 B^2 r^4 - 2 A^2 r^4 (3 + 2 B r^2) + A r^2
(3 + 16 B r^2 + 2 B^2 r^4) + 8 n^2 (2 + A (r^2\\\nonumber&& + B
r^4)) + 2 n (-11 + 2 B r^2 - 2 B^2 r^4 - A r^2 (5 + 12 B r^2 + B^2
r^4) + 2 A^2 (r^4 + B r^6)))\\\nonumber&& + e^{A r^2} (-4 A^2 r^4 (3
+ 5 B r^2) + 2 (5 - 10 B r^2 + 8 B^2 r^4) + A r^2 (19 + 56 B r^2 +
16 B^2 r^4\\\nonumber&& + 4 B^3 r^6) + 4 n^2 (6 + r^4 (A + A B
r^2)^2 + A (6 r^2 + 8 B r^4)) -2 n (15 - 14 B r^2 + 6 B^2 r^4
\\\nonumber&&+A^2 r^4 (-1 - 2 B r^2 + 2 B^2 r^4)+2 A r^2 (10 + 21 B
r^2 + 4 B^2 r^4 + B^3 r^6)))+e^{\frac{Ar^2}{2}} (-5\\\nonumber&& +
28 B r^2 - 16 B^2 r^4 + 2 A^2 r^4 (3 + 8 B r^2 + 4 B^2 r^4)-A r^2
(21 + 58 B r^2 + 30 B^2 r^4
\\\nonumber&&+ 12 B^3 r^6)-8 n^2 (2 + A r^2 (3 + 5 B r^2) +A^2 (r^4
+ 3 B r^6 + 2 B^2 r^8))+2 n (9 - 18 B r^2\\\label{30}&& + 6 B^2 r^4
+A r^2 (21 + 45 B r^2 + 13 B^2 r^4 + 6 B^3 r^6)+4 A^2 (r^4 + 3 B r^6
+ 2 B^2 r^8))))).
\end{eqnarray}
In \cite{31}, Herrera developed a new technique to explore the
potentially unstable matter configuration and introduced the concept
of cracking. One can analyze the potentially stable and unstable
regions regions depending on the difference of sound speeds, the
region for which radial sound speed is greater than the transverse
sound speed is said to be potentially stable. In Figure 11 and 12,
we plot the radial and transverse speeds for different strange star
candidates. Here, the values of $v^2_{sr}$ and $v^2_{st}$ lies in
the range $0<|v^2_{i}|<1$ within the anisotropic matter
configuration. We also present the evolution of $v^2_{st}-v^2_{sr}$
in Figure 13, which shows that difference of two sound speeds
satisfies the inequality $|v^2_{st}-v^2_{sr}|\leq1$. Moreover,
$v^2_{st}-v^2_{sr}$ retain the similar sign for Her X-1 whereas the
sign changes in case of SAX J 1808.4-3658 and 4U 1820-30. Hence, our
proposed strange star model is stable for Her X-1 and unstable for
both SAX J 1808.4-3658 and 4U 1820-30.
\begin{figure}
\centering \epsfig{file=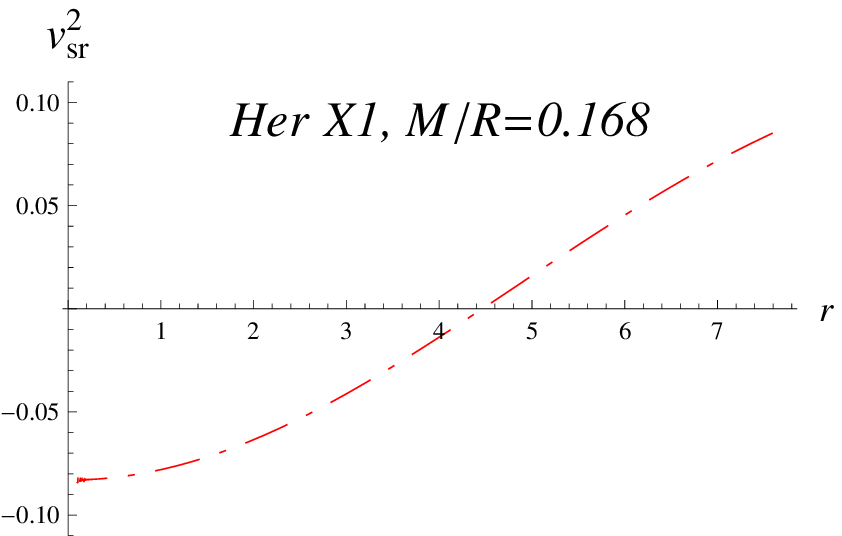, width=.34\linewidth,
height=1.3in}\epsfig{file=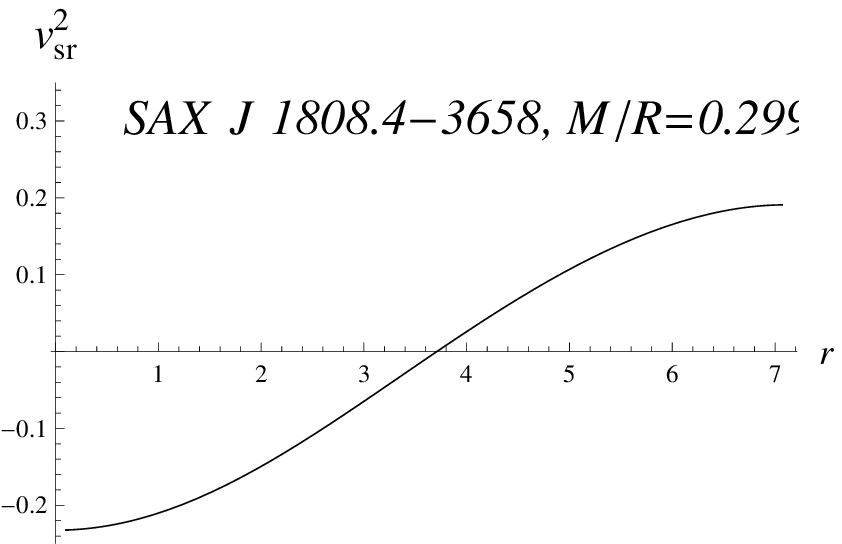, width=.36\linewidth,
height=1.3in}\epsfig{file=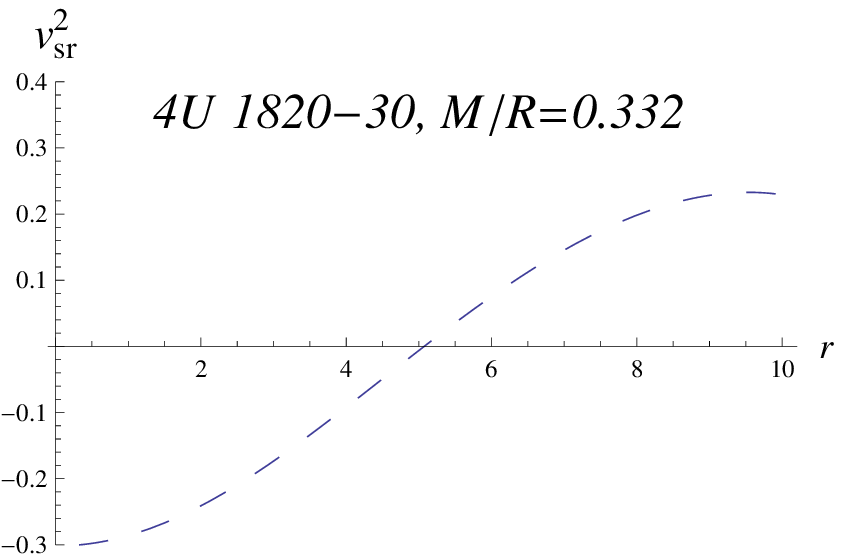, width=.34\linewidth,
height=1.3in}\caption{Evolution of $v^2_{sr}$ versus radial
coordinate $r(km)$ at the stellar interior of strange star
candidates.}
\end{figure}\\
\begin{figure}
\centering \epsfig{file=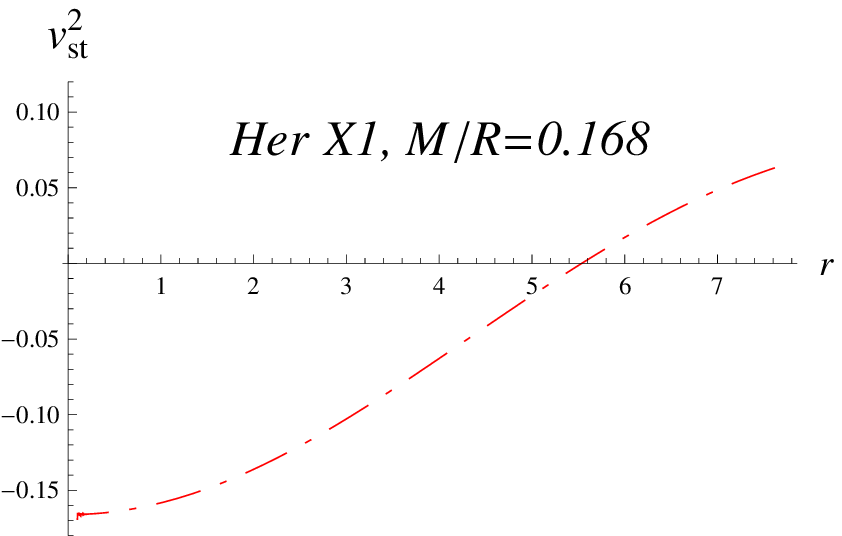, width=.34\linewidth,
height=1.3in}\epsfig{file=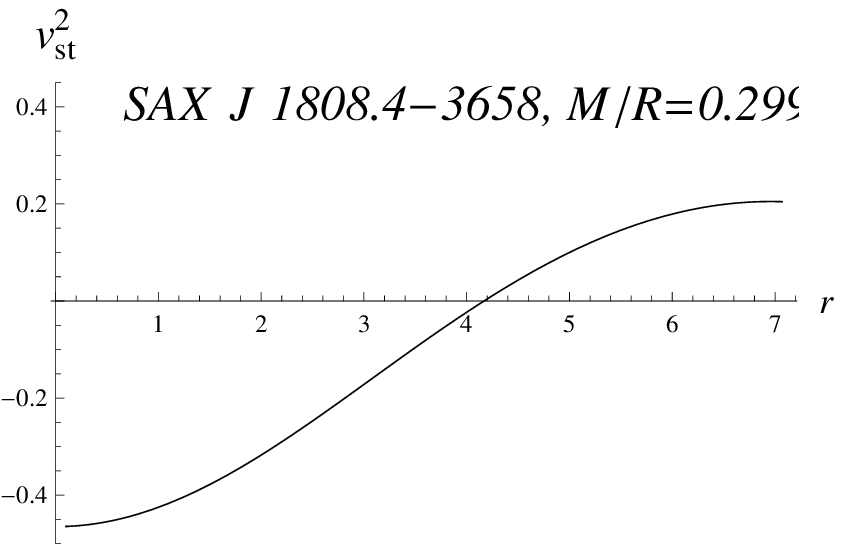, width=.36\linewidth,
height=1.3in}\epsfig{file=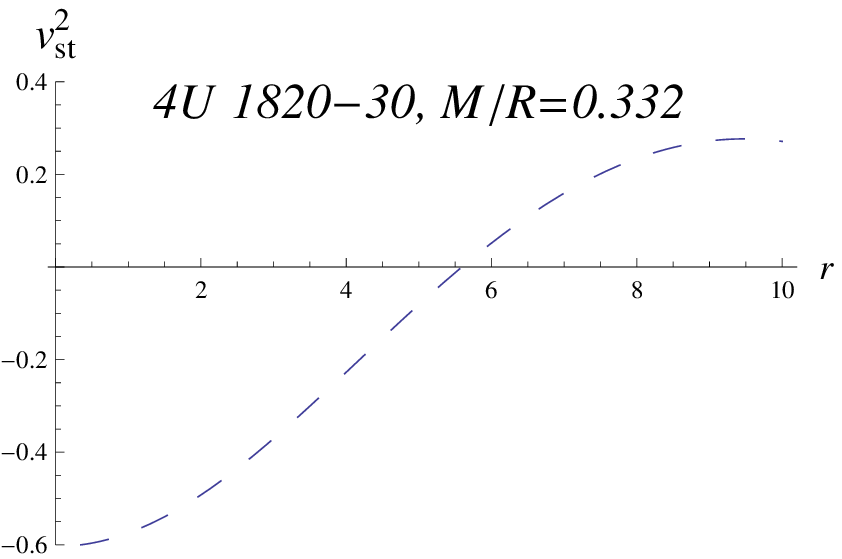, width=.34\linewidth,
height=1.3in}\caption{Evolution of $v^2_{st}$ versus radial
coordinate $r(km)$ at the stellar interior of strange star
candidates.}
\end{figure}\\
\begin{figure}
\centering \epsfig{file=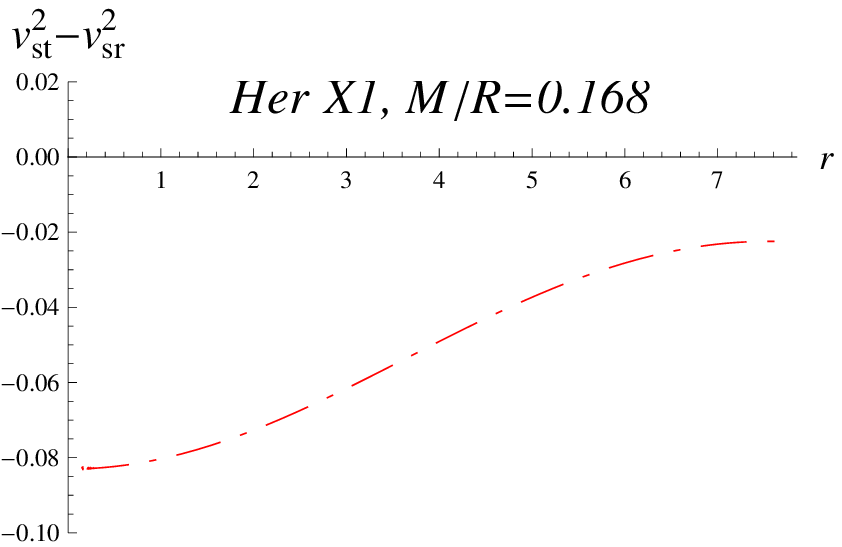, width=.34\linewidth,
height=1.3in}\epsfig{file=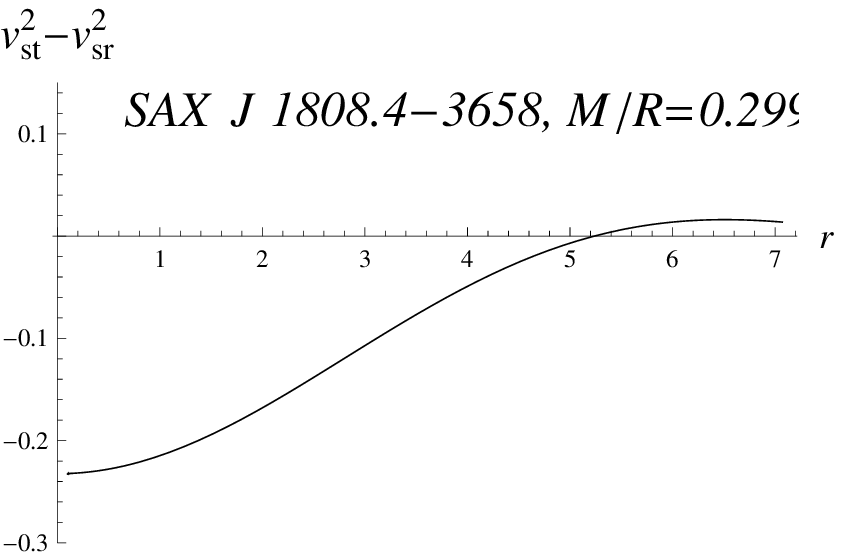, width=.36\linewidth,
height=1.3in}\epsfig{file=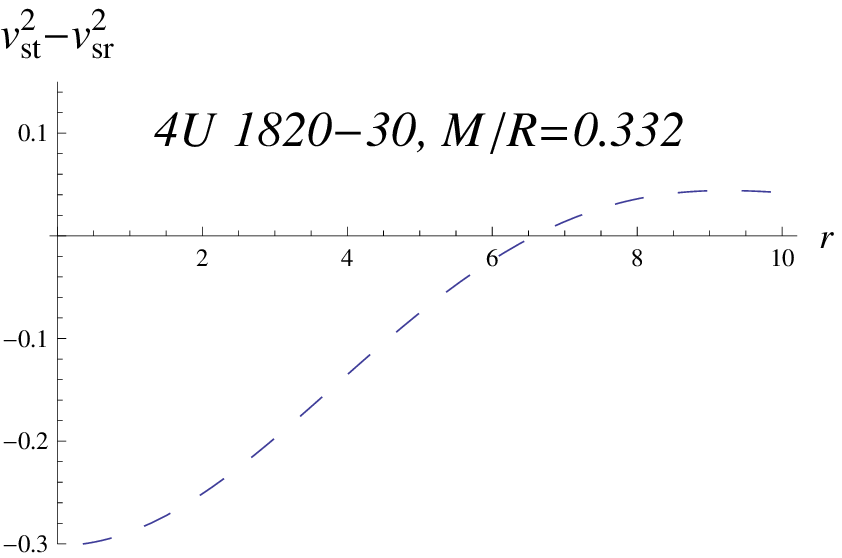, width=.34\linewidth,
height=1.3in}\caption{Evolution of $v^2_{st}-v^2_{sr}$ versus radial
coordinate $r(km)$ at the stellar interior of strange star
candidates.}
\end{figure}
\subsection{Surface Redshift}

The compactness of star in $f(T)$ gravity can be found as

\begin{eqnarray}\nonumber&&
u=\frac{M}{b}=\frac{e^{-2 A R^2} \pi R \beta }{9 (A
R^2)^{\frac{3}{2}}}\left[-36 (20 A^2 + 2 A B + B^2)e^{2 A R^2}
\sqrt{\pi}R^2 Erf[\sqrt{A R^2}]\right.\\\nonumber&&\left.+16(24A^2+
A B + B^2) e^{2 A R^2} Sqrt{6 \pi}R^2 Erf[\sqrt{\frac{3}{2}} \sqrt{A
R^2}]-3 (4 \sqrt{ A R^2}(-B^2\right.\\\nonumber&&\left.(3 - 8
e^{\frac{A R^2}{2}} + 6 e^{A R^2})R^2+A (27 - 24 e^{\frac{3A
R^2}{2}} + 3 e^{2 A R^2} + 6 B R^2-8e^{\frac{A R^2}{2}}(9
\right.\\\nonumber&&\left.+ 2 B R^2)+6 e^{A R^2} (11 + 2 B R^2)))-48
A^2 e^{2 A R^2} \sqrt{2 \pi}R^2 Erf[ \frac{\sqrt{A R^2}}{\sqrt{2}}]
+3(32 A^2\right.\\\label{31}&&\left.-2AB+B^2)e^{2AR^2}\sqrt{2\pi}R^2
Erf[\sqrt{2} \sqrt{A R^2}])\right].
\end{eqnarray}

The surface redshift corresponding to compactness is given by

\begin{eqnarray}\nonumber&&
1+Z_s=(1-2u)^{-1/2}=\{1-\frac{2e^{-2 A R^2} \pi R \beta }{9 (A
R^2)^{\frac{3}{2}}}\left[-36 (20 A^2 + 2 A B + B^2)e^{2 A R^2}
\right.\\\nonumber&&\left.\sqrt{\pi}R^2 Erf[\sqrt{A R^2}]+16(24A^2+
A B + B^2) e^{2 A R^2}\sqrt{6 \pi}R^2 Erf[\sqrt{\frac{3}{2}} \sqrt{A
R^2}]\right.\\\nonumber&&\left.-3 (4 \sqrt{ A R^2}(-B^2(3 - 8
e^{\frac{A R^2}{2}} + 6 e^{A R^2})R^2+A (27 - 24 e^{\frac{3A
R^2}{2}} + 3 e^{2 A R^2} + 6 B
R^2\right.\\\nonumber&&\left.-8e^{\frac{A R^2}{2}}(9+ 2 B R^2)+6
e^{A R^2} (11 + 2 B R^2)))-48 A^2 e^{2 A R^2} \sqrt{2 \pi}R^2 Erf[
\frac{\sqrt{A R^2}}{\sqrt{2}}]\right.\\\label{32}&&\left.+3(32
A^2-2AB+B^2)e^{2AR^2}\sqrt{2\pi}R^2 Erf[\sqrt{2} \sqrt{A
R^2}])\right]\}.
\end{eqnarray}

In Figure 13, we show the redshift plot for the star Her X-1 and
maximum redshift turns out to be $Z_s=0.025$.

\begin{figure}
\centering \epsfig{file=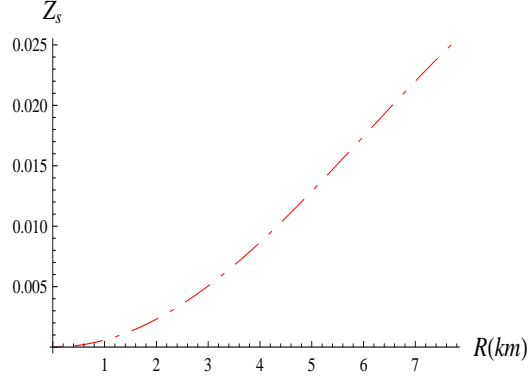, width=.5\linewidth,
height=2in}\caption{Evolution reshift $Z_s$ versus radial coordinate
$r(km)$ at the stellar interior of strange star Her X-1.}
\end{figure}

\section{Conclusion}

In this paper, we have addressed the issue of modeling the
spherically anisotropic compact stars in $f(T)$ gravity using the
off diagonal tetrad field. For the analytic solution of the system
of differential equations, we have assumed the KB form of the
metyric components and the power law form of $f(T)$ model, as
$f(T)=\beta T^n$, with $n\neq1$ and $\beta$ is an arbitrary small
constant, which plays the role of departure of these theories from
GR. This approach leads to the set field equations, which second
order nonlinear differential equations, for the analytic form of
solutions we assume that the metric coefficients for the interior
and exterior regions of a star continuous over the boundary. This is
the smooth matching of the general spherical interior geometry with
the Schwarzschild exterior geometry, because in both regions metric
function are continuous upto fist order of derivatives (GR and
$f(T)$ are second order derivative theories). This matching helps to
use the observational data of stars for the calculation of arbitrary
constants appearing in the system due to parameterization of metric
functions.

We have determined the explicit form of the matter density and
pressure components, which are regular inside the stars as shown in
figures ({\textbf{1-3}}). The behavior of EoS parameter implies that
near the boundary the matter inside the stars behaves as
quintessence in the presence of $f(T)$ terms (see figure
\textbf{6}). The anisotropic parameter $\Delta=0$ at center vanished
and it becomes $\Delta>0$, everywhere inside the star except at
center, which predicts that there may exist the repulsive force
which leads to the formation of more massive star. The physical
viability of the solutions lies in the fact that solutions satisfies
the energy conditions (Figure \textbf{9}). The analysis of TOV
equation presented in this paper shows that the static equilibrium
can be achieved for strange stars in $f(T)$ as shown in
figure\textbf{ 10}. The stability of the proposed model is given in
figures \textbf{11-13}, on the physical investigations of these
figures predict that our proposed strange star model is stable for
Her X-1, and unstable for both SAX J 1808.4-3658 and 4U 1820-30. The
maximum redshift of Her X-1 turn out be 0.025 which is shown in
Figure \textbf{14}.

\renewcommand{\theequation}{A.\arabic{equation}}
\setcounter{equation}{0}
\section*{Appendix A}

\begin{eqnarray}
\nonumber&& \frac{\partial\rho}{\partial
r}=\frac{n2^{n-1}\beta}{r^{2n+1}}\left\{(e^{-Ar^2})^n\left(e^{\frac{Ar^2}{2}}-1\right)^{n-2}
\left(e^{\frac{Ar^2}{2}}-1-2Br^2\right)^{n-3}\right\}\left[7Ar^2-e^{\frac{5Ar^2}{2}}\right.
\\\nonumber&&\left.+1-10Br^2+18ABr^4+12AB^2r^6+8AB^3r^8+e^{2Ar^2}(5+4n^2-2Ar^2+
6Br^2\right.\\\nonumber&&\left.
+2n(-3+Ar^2-Br^2))+4n^2(1+A(r^2+2Br^4))^2-2n(2B^2r^4-7Br^2+2\right.
\\\nonumber&&\left.+3A^2r^4(1+2Br^2)^2+Ar^2(7
+15Br^2+4B^2r^4+4B^3r^6))-e^{\frac{3Ar^2}{2}}(10+4Br^2\right.
\\\nonumber&&\left.+8B^2r^4-2A^2r^4(3+2Br^2)+Ar^2(3+16Br^2+2B^2r^4)
+8n^2(2+A(r^2+Br^4))\right.
\\\nonumber&&\left.+2n(2Br^2-11-2B^2r^4)-Ar^2(5+12Br^2+B^2r^4)+2A^2r^4(1+Br^2)))+e^{Ar^2}\right.
\\\nonumber&&\left.\times(10-20Br^2+16B^2r^4-4A^2r^4(3+5Br^2)+Ar^2(56Br^2+16B^2r^4+4B^3r^6
\right.
\\\nonumber&&\left.+19)+4n^2
(6+r^4(A+ABr^2)^2+A(6r^2+8Br^4))-2 n (15 - 14 B r^2+ 6 B^2 r^4
\right.
\\\nonumber&&\left. +A^2 r^4(-1- 2 B r^2 + 2 B^2 r^4) +
2 A r^2 (10 + 21 B r^2 + 4 B^2 r^4+ B^3 r^6))) +
   e^{\frac{Ar^2}{2}}(-5\right.
\\\nonumber&&\left.+ 28 B r^2 - 16 B^2 r^4 +2 A^2 r^4 (3 + 8 B r^2 + 4 B^2 r^4) -
A r^2 (58 B r^2 + 30 B^2 r^4 + 12 B^3 r^6\right.
\\\nonumber&&\left.+21) - 8 n^2 (2 + A r^2 (3 + 5 B r^2)+
A^2 (r^4 + 3 B r^6 + 2 B^2 r^8)) +2 n (-18 B r^2+9+\right.
\\\label{01}&&\left. 6 B^2 r^4 + A r^2 (21 + 45 B r^2
+ 13 B^2 r^4 + 6 B^3 r^6) + 4 A^2 (r^4 + 3 B r^6 + 2 B^2
r^8)))\right],
\end{eqnarray}
\begin{eqnarray}\nonumber
&&\frac{\partial p_r}{\partial
r}=\frac{n2^{n-1}\beta}{r^{2n+1}}\left\{(e^{-Ar^2})^n(\left(e^{\frac{Ar^2}{2}}-1\right)
\left(e^{\frac{Ar^2}{2}}-1-2Br^2\right))^{n-2}\right\}\left[
 2 e^{\frac{3 A r^2}{2}}(-2 + n)\right.
\\\nonumber&\times&\left.(1 + B r^2)+e^{2 A r^2}-(2n-1) (1 + 2 B r^2) (1 +
    A (r^2 + 2 B r^4)) +
 e^{A r^2} (6 + 10 B r^2 \right.
\\\nonumber&+&\left. A (r^2 + 2 B r^4 + 2 B^2 r^6) -
    2 n (3 + 4 B r^2 + A (r + B r^3)^2)) +
 2e^{\frac{ A r^2}{2}}(-(1 + 2 B r^2) (2\right.
\\\label{02}&+&\left.A (r^2 + B r^4)) +
    n (3 + 5 B r^2 + 2 A (r^2 + 3 B r^4 + 2 B^2 r^6)))\right],
\end{eqnarray}
\begin{eqnarray}
\nonumber&& \frac{\partial p_t}{\partial
r}=\frac{-n2^{n-1}\beta}{r^{2n+1}}\left\{(e^{-Ar^2})^n(\left(e^{\frac{Ar^2}{2}}-1\right)
\left(e^{\frac{Ar^2}{2}}-1-2Br^2\right))^{n-3}\right\}\left[e^{3Ar^2}(n-1)-1\right.
\\\nonumber&&\left.-2 A r^2-10 AB r^4-6 B^2 r^4 -
 18A B^2 r^6-6 B^3 r^6-12 A B^3 r^8-4 B^4 r^8+e^{\frac{5Ar^2}{2}}(n-1)\right.
\\\nonumber&&\left.\times(2n-6-B r^2(4+A r^2))-
 2n^2 (1+B r^2)(1 + A (r^2 + 2Br^4))^2 +n(2 B r^2+5 B^2 r^4\right.
\\\nonumber&&\left.+3+ 2 B^3 r^6 +r^4 (1 + B r^2) (A + 2 A B r^2)^2 +A r^2 (6 + 25 B r^2+ 39 B^2 r^4 + 28 B^3 r^6 \right.\\\nonumber&&\left.+ 4 B^4 r^8)) - e^{2 A
r^2}(15 + 16 B r^2 + 6 B^2 r^4 +
 A r^2 (2 + 10 B r^2 + 3 B^2 r^4)- A^2r^4(1+B r^2)\right.
\\\nonumber&&\left. +2 n^2 (5 + B r^2 + 2 A (r^2 + B r^4))+n (-25 - 18 B r^2 - 5 B^2 r^4 -
A r^2 (17 B r^2 + 4 B^2 r^4\right.
\\\nonumber&&\left.+6) + 2 A^2 (r^4 + B r^6))) +
e^{\frac{3Ar^2}{2}}(20 + 24 B r^2 + 24 B^2 r^4 + 6 B^3 r^6 - 3 A^2
(r^4 + 2 B r^6)\right.
\\\nonumber&&\left. +
A r^2 (8 + 34 B r^2 + 15 B^2 r^4 + B^3 r^6) + 2 n^2 (10 + 4 B r^2 +
r^4 (A + A B r^2)^2 + 2 A r^2 (4\right.
\\\nonumber&&\left. + 6 B r^2 + B^2 r^4)) + n (A^2 r^4 (5 + 10 B r^2 + 2 B^2 r^4) -
2 (20 + 16 B r^2 + 10 B^2 r^4 + B^3 r^6)\right.
\\\nonumber&&\left. -A r^2 (24 + 70 B r^2 + 35 B^2 r^4 + 5 B^3 r^6))) -
 e^{A r^2} (16 B r^2 + 36 B^2 r^4 + 18 B^3 r^6 + 4 B^4 r^8\right.
\\\nonumber&&\left.+15 -
 3 A^2 (r^4 + 3 B r^6 + 2 B^2 r^8) +A r^2 (12 + 52 B r^2 + 39 B^2 r^4 + 6 B^3 r^6 - 2 B^4 r^8)\right.
\\\nonumber&&\left. +2 n^2 (10 + 6 B r^2 + 4 A r^2 (3 + 6 B r^2 + 2 B^2 r^4) +
A^2 r^4 (3 + 9 B r^2 + 7 B^2 r^4 + B^3 r^6))\right.
\\\nonumber&&\left. +n (-35 - 28 B r^2 - 30 B^2 r^4 - 6 B^3 r^6 +
A^2 r^4 (3 + 9 B r^2 + 8 B^2 r^4 + 2 B^3 r^6) -A r^2 (36\right.
\\\nonumber&&\left. + 118 B r^2 + 97 B^2 r^4 + 30 B^3 r^6 +2 B^4 r^8))) +
 e^{\frac{A r^2}{2}} (4 B r^2 + 24 B^2 r^4 + 18 B^3 r^6 + 8 B^4 r^8\right.
\\\nonumber&&\left.+6 + A r^2 (8 + 37 B r^2 + 45 B^2 r^4 + 17 B^3 r^6 - 2 B^4 r^8) -
A^2r^4(4 B r^2 + 6 B^2 r^4 + 4 B^3 r^6\right.
\\\nonumber&&\left.+1) +2 n^2 (5 + 4 B r^2 + 2 A r^2 (4 + 10 B r^2 + 5 B^2 r^4) +
A^2 r^4 (12 B r^2 + 14 B^2 r^4 + 4 B^3 r^6\right.
\\\nonumber&&\left.+3)) + n (-2 (8 + 6 B r^2 + 10 B^2 r^4 + 3 B^3 r^6) +
 A^2 r^4 (-1 - 4 B r^2 - 2 B^2 r^4 + 4 B^3 r^6)\right.
\\\label{03}&&\left. - A r^2 (24 + 89 B r^2 + 105 B^2 r^4 + 53 B^3 r^6 + 6 B^4 r^8)))\right]
\end{eqnarray}

\end{document}